\documentstyle[12pt,epsf,times]{article}  

\begin{document}
\begin{titlepage}
\def\thefootnote{\fnsymbol{footnote}}

\begin{flushleft}
\setlength{\baselineskip}{13pt}
ITP-UH-16/95 \hfill May 1995\\
cond-mat/9505146
\end{flushleft}

\vspace*{\fill}
\begin{center}
{\Large On the Hubbard Model in the Limit of Vanishing Interaction} \\
\vfill
\vspace{1.5 em}
{\sc Holger Frahm}\footnote{e-mail: {\tt frahm@itp.uni-hannover.de}}
and
{\sc Markus P.\ Pfannm\"uller}\footnote
               {e-mail: {\tt pfannm@itp.uni-hannover.de}}\\
{\sl Institut f\"ur Theoretische Physik, Universit\"at Hannover\\
D-30167~Hannover, Germany}\\
\vfill
ABSTRACT
\end{center}
\begin{quote}
We address the question how a correspondence between the particle like
excitations in the one dimensional Hubbard model (i.e.\ ``holons'' and
``spinons'') and the free fermionic picture can be estabilished in the
limit of vanishing interaction by studying the finite size spectrum in the
framework of the Bethe Ansatz.  Special attention has to be paid to the
case of a vanishing magnetic field where the two bands of excitations in
either description are degenerate.  The interaction lifts this degeneracy.
\end{quote}
\vfill
PACS-numbers:	71.27.+a~\      
		75.10.Lp~\      
		05.70.Jk~\      

\vfill
\setcounter{footnote}{0}
\end{titlepage}

The Hubbard model is one of the most studied models for interacting
electrons on a one-dimensional lattice. Following Lieb and Wu's
\cite{liwu:68} Bethe Ansatz (BA) solution many exact results have
been obtained. These provide detailed understanding of the
thermodynamics \cite{taka:72}, excitation spectrum
\cite{ovch:70}\nocite{coll:74}--\cite{esko:94}, finite size corrections
\cite{woec:87b,woyn:89} and asymptotics of correlation functions
\cite{frko:90,frko:91} which are believed to show the generic behaviour of
systems of interacting electrons in one spatial dimension.

The Hamiltonian is given in terms of standard fermionic creation resp.\
annihilation operators $\Psi^{\dagger}_{j,\sigma}$ and $\Psi_{j,\sigma}$ of
electrons with spin $\sigma$ at site $j$ and the corresponding occupation
numbers $n_{j,\sigma} = \Psi^{\dagger}_{j,\sigma}\Psi_{j,\sigma}$
\begin{equation}
{\cal H}  =  \sum_{j=1}^{N}\Biggl[\,\sum_{\sigma=\uparrow,\downarrow}\left(
        \Psi^{\dagger}_{j+1,\sigma}\Psi_{j,\sigma}+
        \Psi^{\dagger}_{j,\sigma}\Psi_{j+1,\sigma} \right)
      +4 u \, n_{j,\uparrow}n_{j,\downarrow}
      + \mu\, (n_{j,\uparrow}+n_{j,\downarrow})
      - \frac{h}{2}\, (n_{j,\uparrow}-n_{j,\downarrow})\Biggr]
\label{H}
\end{equation}
where $u$ is the strength of the on-site Coulomb repulsion, $\mu$ the
chemical potential and $h$ an additional external magnetic field. In the
following we consider the repulsive case $u\ge 0$ with $h\ge 0\,$.

For vanishing coupling constant $u=0$ the model simply describes two
independent generations of free fermions. This fact allows, of course, to
extract the physical properties of the system in a much simpler way than
through the BA. Nevertheless, there have been studies of the BA solution in
the $u\to0$ limit, mainly motivated by the desire to have a reliable test
for perturbative schemes expanding around the free fermionic limit. These
studies have mainly concentrated on the $u$-dependence of the ground state
energy for the half-filled band, finding that the ground state energy can
be expanded in an asymptotic series in $u$ which is reproduced correctly by
standard perturbation theory \cite{taka:71}\nocite{ecpo:79}--\cite{mevo:89}.

In this letter we extend the study of the $u\to 0$ limit to include the
behaviour of the low-lying excitations. In particular, we address the
question how a correspondence between the particle like excitations of the
interacting system (i.e.\ ``holons'' and ``spinons'') as obtained from the
BA and those present in the free fermion spectrum can be established in
this limit.
It turns out that in a finite magnetic field there is a one to one
correspondence between the excitations in either description.
In absence of a magnetic field the single particle energies of {\em
non\/}interacting spin-$\uparrow$ and spin-$\downarrow$ electrons are
identical. This is reproduced by the BA solutions where the Fermi
velocities of spin- and charge excitations are degenerate in the limit
$u\to0\,$. As a consequence {\em both} pictures lead to a correct description
of the spectrum of (at least low-lying) excitations, charge an spin degrees
of freedom can be assigned to the two degenerate bands of excitations in an
almost arbitrary way. For finite values of $u$ this degeneracy is lifted.
We investigate how the difference between the two velocities develops as a
function of the interaction strength $u$ for zero magnetic field.

We recall that the density of charge and spin waves in the thermodynamic
limit is given in terms of an inhomogeneous integral equation
\begin{equation}
\rho = \rho^{(0)}+\mbox{\boldmath$\hat{K}$}*\rho
\label{Irho}
\end{equation}
where $\rho$ and $\rho^{(0)}$ are column vectors with entries
\begin{equation}
\rho={\rho_c(k) \choose \rho_s(\lambda)} ,\hspace{1cm}
\rho^{(0)}= {\frac{1}{2\pi} \choose 0}
\end{equation}
and {\boldmath $\hat{K}$} is a $2\times2$ matrix
whose elements are integral operators, namely
\begin{equation}
\mbox{\boldmath$\hat{K}$}=\left(\begin{array}{cc}
     0 & \cos k\int\limits_{-\lambda_0}^{+\lambda_0}
                d\lambda'K_1(\sin k -\lambda') * \rule[-8mm]{0pt}{8mm}\\
    \int\limits_{-k_0}^{+k_0}dk' K_1(\lambda-\sin k') * &
   -\int\limits_{-\lambda_0}^{+\lambda_0}d\lambda'K_2(\lambda -\lambda') *
 \end{array}\right)\ .
\label{Kh}
\end{equation}
The renormalized energies of the corresponding excitations read
\begin{equation}
\epsilon = \epsilon^{(0)}+\mbox{\boldmath$\hat{K}$}^{T}*\epsilon
\label{Ieps}
\end{equation}
with
\begin{equation}
\epsilon={\epsilon_c(k) \choose \epsilon_s(\lambda)}\ ,\hspace{1cm}
\epsilon^{(0)}={\mu-\frac{h}{2}-2 \cos k \choose h}\ .
\end{equation}
The integral operator matrix $\mbox{\boldmath$\hat{K}$}^T$ is the transpose of
{\boldmath$\hat{K}$}, namely
\begin{equation}
\mbox{\boldmath$\hat{K}$}^T=\left(\begin{array}{cc}
     0 & \int\limits_{-\lambda_0}^{+\lambda_0}d\lambda'K_1(\sin k
	-\lambda') *
          \rule[-8mm]{0pt}{8mm}\\
    \int\limits_{-k_0}^{+k_0}dk'\cos k'K_1(\lambda-\sin k') * &
   -\int\limits_{-\lambda_0}^{+\lambda_0}d\lambda'K_2(\lambda -\lambda') *
 \end{array}\right)\ .
\label{Kht}
\end{equation}
In these equations the kernels $K_1$ and $K_2$ are given by
\begin{equation}
   K_1(x)= \frac{1}{2\pi}\ \frac{2u}{u^2+x^2}\ ,\hspace{1cm}
   K_2(x)= \frac{1}{2\pi}\ \frac{4u}{4u^2+x^2}\ .
\label{kernel}
\end{equation}

For a comparison to the free fermionic description the finite size
corrections of the BA spectrum as calculated in \cite{woyn:89} are
particularly useful.
The energies and momenta of the low lying excitations are given by
\begin{eqnarray}
E - E_0 &  = & \frac{2\pi}{N} \Bigl[
          \frac{1}{4} \Delta N^{T}(\mbox{\boldmath$Z$}^{-1})^{T}
           \mbox{\boldmath$VZ$}^{-1}\Delta N
         + D^{T}\mbox{\boldmath$ZVZ$}^{T}D\nonumber\\
        &    &+ v_c (N^+_c + N^-_c)
         +v_s (N^+_s + N^-_s)\Bigr] + o\left(\frac{1}{N}\right)\ ,
\label{BAex}\\
P - P_0 & = & \frac{2\pi}{N} (
        \Delta N^{T}D + N^+_c - N^-_c + N^+_s - N^-_s)\nonumber\\
&  &     +2D_ck_{F,\uparrow}+2(D_c+D_s)k_{F,\downarrow}\ .
\label{BAp}
\end{eqnarray}
Here {\boldmath$V$} denotes the diagonal matrix $\mbox{\boldmath
$V$}=\mbox{diag}(v_c,v_s)\,$ of the Fermi velocities of charge and spin waves
\begin{equation}
   v_c = \frac{\epsilon'_c(k_0)}{2\pi\rho_c(k_0)}\ ,\hspace{1cm}
   v_s = \frac{\epsilon'_s(\lambda_0)}{2\pi\rho_s(\lambda_0)}\ .
\label{vv}
\end{equation}
The matrix
\begin{equation}
\mbox{\boldmath $Z$} = \left(\begin{array}{cc} Z_{cc} & Z_{cs} \\
                             Z_{sc} & Z_{ss} \end{array}
    \right) =
     \left(\begin{array}{cc} \xi_{cc}(k_0) & \xi_{sc}(k_0) \\
                             \xi_{cs}(\lambda_0) & \xi_{ss}(\lambda_0)
          \end{array}\right)^T
\label{Z}
\end{equation}
is given in terms of the dressed charge matrix {\boldmath $\xi$}
which is defined by the integral equation
\begin{equation}
\mbox{\boldmath $\xi$} =  \mbox{\boldmath $I$}
  +\mbox{\boldmath$\hat{K}$}^{T}*\mbox{\boldmath $\xi$}
\label{Ixi}
\end{equation}
where {\boldmath $I$} is the $2\times2$ unit matrix.
The vectors
\begin{equation}
\Delta N = {\Delta N_c \choose \Delta N_s}\ ,\hspace{2cm}
D = {D_c \choose D_s}
\label{DN_D}
\end{equation}
and the positive integers $N^{\pm}_c$ and $N^{\pm}_s$ characterize the
excited state. Here $\Delta N_c=\Delta N_\uparrow+\Delta N_\downarrow\,$ and
$\Delta N_s=\Delta N_\downarrow\,$ are related to the change in particle
numbers with respect to their ground state values thus determining charge
and spin of the excitated state, respectively. $D_c= D_\uparrow\,$ and
$D_s=D_\downarrow-D_\uparrow\,$ are given by the number of particles moved
from the left to right Fermi points at $\pm k_{F,\sigma}=\pm \pi n_\sigma$
($n_{\sigma}$ are the total densities of electrons with spin $\sigma$).
Their values are integers or half integers subject to the conditions $D_c
\equiv (\Delta N_c+\Delta N_s)/2\,$ and $D_s\equiv \Delta N_c/2\,$ modulo $1$.
The values of $N^{\pm}_{c,s}$ are the quantum numbers of particle--hole
excitations at the right, resp.\ left Fermi points.

For vanishing $u$ the kernels (\ref{kernel}) become $\delta$-functions and
the solution of Eqs.\ (\ref{Irho}), (\ref{Ieps}), (\ref{Ixi}) is
trivial. However, to determine the Fermi velocities (\ref{vv}) and the
matrix {\boldmath $Z$} (\ref{Z}) these solutions have to be taken at the
boundaries $k_0\,$ and $\lambda_0\,$.
For $\sin k_0 \leq \lambda_0\,$ the solutions
are discontinuous at these points and the limit $u\to0$ has to be performed
{\em after} solving the integral equations.

To see whether this situation can arise we restrict ourselves to
$\lambda_0< \sin k_0$ first. In this case the discontinuities are moved
away from the boundaries entering (\ref{vv}) and (\ref{Z}) and we find
\begin{eqnarray}
   \rho_c(k)&=&\left\{\begin{array}{ccl}
   \frac{1}{\pi} & \mbox{ if } & 0 \leq |k| \leq \arcsin \lambda_0\\
   \frac{1}{2\pi}& \mbox{ if } & \arcsin \lambda_0 < |k|
   \end{array}\right.\ ,
\label{rhocFF}\\
   \rho_s(\lambda)&=&\left\{\begin{array}{ccl}
   \frac{1}{2\pi\cos(\arcsin \lambda)}
   & \mbox{ if } & 0 \leq |\lambda| \leq \sin k_0\\
   0 & \mbox{ if } & \sin k_0 < |\lambda|
   \end{array}\right.\ .
\label{rhosFF}
\end{eqnarray}
(Alternatively one can express the density $\rho_s$ as a function of
quasimomenta $k=\arcsin\lambda\,$ rather than the rapidities $\lambda$
themselves, (\ref{rhosFF}) simplifies to
$\rho_s(k)=\theta(k_0-|k|)/(2\pi)$.)  From these equations we obtain for
the total densities of the charge and spin excitations corresponding to
this state
\begin{equation}
  n_c = \int_{-k_0}^{+k_0}\rho_c(k)\,dk\;
      =\; \frac{k_0}{\pi}+\frac{\arcsin\lambda_0}{\pi}\ , \qquad
  n_s = \int_{-\lambda_0}^{+\lambda_0}
          \rho_s(\lambda)\,d\lambda\;
      =\; \frac{\arcsin \lambda_0}{\pi}
\end{equation}
which allows for the identification of $k_0\,$ and $\lambda_0\,$ in terms of
the Fermi momenta through $k_0=k_{F,\uparrow}\,$ and $\arcsin
\lambda_0=k_{F,\downarrow}\,$.  Thus we find that the condition
$\lambda_0<\sin k_0\,$ is satisfied for {\em any} $h>0$. The case of a
vanishing magnetic field has to treated separately.

The dressed energies are given by
\begin{eqnarray}
   \epsilon_c(k)&=&\left\{\begin{array}{ccl}
   2\mu-4\cos k & \mbox{ if } & 0 \leq |k| \leq \arcsin \lambda_0\\
   \mu-\frac{h}{2}-2\cos k& \mbox{ if } & \arcsin \lambda_0 < |k|
   \end{array}\right.\ ,
\label{epscFF}\\
   \epsilon_s(k)&=&\left\{\begin{array}{ccl}
   \mu+\frac{h}{2}-2\cos k
   & \mbox{ if } & 0 \leq |k| \leq k_0\\
    h & \mbox{ if } & k_0 < |k|
   \end{array}\right.\ .
\label{epssFF}
\end{eqnarray}
{}From Eq.\ (\ref{vv}) we find $v_c = 2\sin(k_{F,\uparrow})$ and $v_s =
2\sin(k_{F,\downarrow})\,$.
Similarly the result for the dressed charge matrix gives
\begin{equation}
\mbox{\boldmath $Z$} = \left(\begin{array}{cc} 1 & 1 \\
                            0 & 1 \end{array} \right)\ .
\label{Zff}
\end{equation}
Now comparing the finite size corrections for the excited states
(\ref{BAex}) with this expression for the matrix {\boldmath $Z$} and the
corresponding free fermion result the two are found to agree.

The case of a vanishing magnetic field $h=0$ needs a special treatment.
In this case we have $\lambda_0=\infty$ and the dressed charge matrix
can be expressed in terms of a single quantitity $\xi$ \cite{woyn:89}
\begin{equation}
  \mbox{\boldmath $\xi$} = \left(\begin{array}{cc}
          \xi(z) & 0 \\
        \frac{1}{2}\xi(z) &\frac{1}{\sqrt{2}}
    \end{array} \right)
\end{equation}
satisfying the integral equation
\begin{equation}
   \xi(z)=1+\int^{z_0}_{-z_0} K(z-z')\xi(z')dz'
\label{xi}
\end{equation}
with the kernel
\begin{equation}
   K(x) =\frac{1}{2\pi} \int_0^{\infty}
	\frac{e^{-\omega}}{\cosh(\omega)}\cos(\omega x) d\omega
\label{KK}
\end{equation}
and $z=\sin k/u\,$. For large $z_0=\sin k_0/u$ the quantity $\xi(z_0)$
entering (\ref{Z}) can be obtained using a perturbative scheme based on
the Wiener--Hopf method \cite{yaya:66}. The result to order $1/z_0$ reads
\cite{frko:90}
\begin{equation}
\xi(z_0) = \sqrt{2}\left(1-\frac{1}{2\pi z_0}\right)\ .
\end{equation}
In the limit $u \rightarrow 0$ we find the following dressed charge matrix
\begin{equation}
\mbox{\boldmath $Z$} = \left(\begin{array}{cc}
          \sqrt{2} & 0 \\
        \frac{1}{2}{\sqrt{2}} &\frac{1}{2}{\sqrt{2}}
    \end{array} \right)\ .
\label{Zffh0}
\end{equation}
One might have expected that the result (\ref{Zff}) for {\boldmath $Z$}
holds even for a vanishing magnetic field as there is no dependence on
$h$. This is indeed the case. ``Holons'' and ``spinons'' are certain
combinations of spin-$\uparrow$ and spin-$\downarrow$ electrons. For
vanishing magnetic field these combinations become arbitrary since
spin-$\uparrow$ and spin-$\downarrow$ electrons have equal energies. The
Fermi velocities are equal, $v_c=v_s=2\sin k_F$, and thus the matrix
{\boldmath $V$} is proportional to the unit matrix,  $\mbox{\boldmath $V$} =
\,2\sin (k_F) {1\,0\choose0\,1}\,$.  Of physical relevance are only the
combinations $(\mbox{\boldmath $Z$}^{-1})^T\mbox{\boldmath $VZ$}^{-1}$ and
$\mbox{\boldmath $ZVZ$}^{T}$ which enter expression (\ref{BAex}) for the
excited states. For both choices of {\boldmath $Z$} the results coincide.

\begin{figure}[h]
\begin{center}
\leavevmode
\epsfxsize=\textwidth
\epsfbox{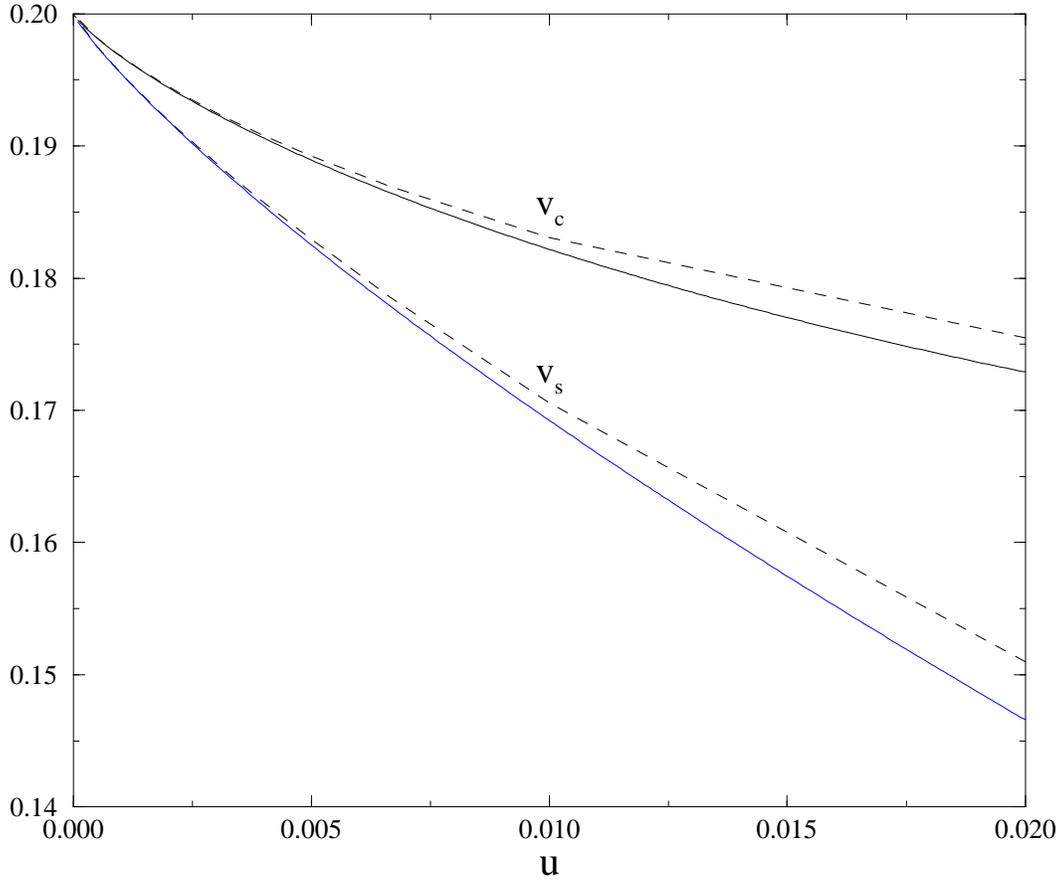}
\vspace{-1.1cm}
\caption{\label{fig1}
Fermi velocities for $h=0$ and $\sin k_0 = 0.1$ as a function of $u$.
Solid lines correspond to numerical solutions of the integral equations
(\protect\ref{Iect}), dashed lines to the asymptotic expressions
(\protect\ref{vasy0}).}
\end{center}
\end{figure}

The degeneracy of the Fermi velocities of charge and spin wave exciations
is lifted by the interaction. Using (\ref{vv}) the Fermi velocities for
$h=0$ can be expressed as
\begin{equation}
   v_c = {1\over 2\pi}\ \frac{g(z_0)}{f(z_0)}\ ,
   \qquad
   v_s = {1\over 2\pi}\ \frac{\int^{z_0}_{-z_0}e^{\frac{\pi}{2}z}g(z)dz}
           {\int^{z_0}_{-z_0}e^{\frac{\pi}{2}z}f(z)dz}\ .
\label{vv0}
\end{equation}
Here $f(z)$ and $g(z)$ are are the density $\rho_c$ and the derivative of
the dressed energy $\epsilon'_c$ as a function of the variable $z$ given in
terms of the following integral equations (remember that we have $uz=\sin
k<1$)
\begin{eqnarray}
   f(z) & = & \frac{1}{2\pi\sqrt{1- u^2 z^2}}
   + \int^{z_0}_{-z_0} K(z-z')f(z')dz'\ ,
  \nonumber \\
   g(z) & = & \frac{2uz}{\sqrt{1- u^2 z^2}}
   + \int^{z_0}_{-z_0} K(z-z')g(z')dz'
\label{Iect}
\end{eqnarray}
with the kernel $K$ given by Eq.\ (\ref{KK}). Again, the quantities
necessary to compute the Fermi velocities (\ref{vv0}) for small $u$, i.e.\
$z_0\approx\infty\,$, can be obtained from these equations using the
Wiener-Hopf method. A complication is given by the explicit $u$--dependence
of the driving terms. However, for small densities (i.e.\ $uz_0=\sin k_0\ll
1$) they can be expanded up to linear order in $uz$.  For $f$ this results
in Eq.~(\ref{xi}) for the dressed charge $\xi$ (up to a factor of
$1/2\pi$). In the equation for $g$ the driving term is replaced by $2uz$.
For $0 < u\ll\sin(k_0)<1$ we finally get the following result for the Fermi
velocities
\begin{eqnarray}
  v_c & = & 2u\left[z_0 -\frac{1}{\pi}\ln(z_0)
          +\frac{1}{\pi}\ln\left(\frac{2}{\pi}\right)\right]\ ,
\nonumber \\
  v_s & = & 2u\left[z_0 -\frac{1}{\pi}\ln(z_0)
          +\frac{1}{\pi}\ln\left(\frac{2}{\pi}\right)
          -\frac{2}{\pi}\right]\ .
\label{vasy0}
\end{eqnarray}
The leading term $2uz_0\,$ is simply the free fermion result $2\sin
k_0\,$. The logarithmic corrections $\propto u\ln u$ are probably just a
consequence of the expression of the velocities in terms of $z_0\,$ rather
than the electron density $n_c\,$. To prove this analytically the Wiener-Hopf
scheme mentioned above has to be performed to order $z_0^{-2}$ which raises
questions in its quality. However, numerical solution of the integral
equations (\ref{Iect}) suggests the absence of logarithmic corrections in
$v_\alpha(n_c,u)$. In Fig.\ \ref{fig1} we present the Fermi velocities
for a fixed value of $\sin k_0=0.1$
which are computed from the numerical solution of the integral equations
(\ref{Iect}) in comparison with Eqs.\ (\ref{vasy0}).  Because of the various
approximations which were necessary to derive Eqs.~(\ref{vasy0}) we expect
the results only do be correct up to the order of $u$.

\begin{figure}[h]
\begin{center}
\leavevmode
\epsfxsize=\textwidth
\epsfbox{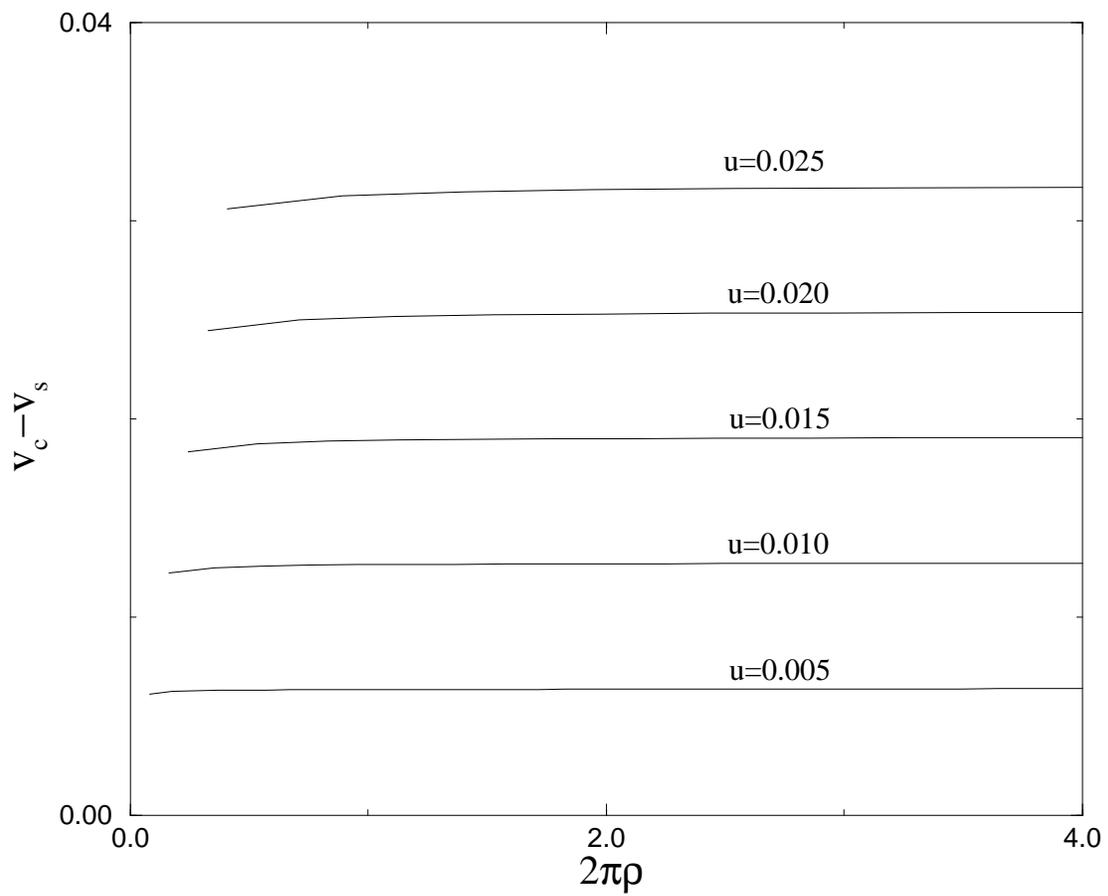}
\vspace{-1.1cm}
\caption{\label{fig2}
Difference of Fermi velocities for $h=0$ as a function of the total density
computed from numerical solutions of Eqs.\ (\protect\ref{Iect}).}
\end{center}
\end{figure}

An interesting observation is that, in leading order, the gap between
charge and spin wave excitations is a linear function of the interaction $u$
\begin{equation}
   v_c-v_s = \frac{4u}{\pi}\ .
\label{gap}
\end{equation}
Fig.\ \ref{fig2} shows the difference of the Fermi velocities as
a function of the total density of particles for various values of $u$.
As for Fig.\ \ref{fig1} the data were computed from numerical solutions
of the integral equations. As long as $\rho$, i.e.\ $\sin k_0\,$, is not
to small, we find exactly the behaviour as predicted by Eq.~(\ref{gap}).
For $\rho\rightarrow 0$ at fixed $u$ one has $z_0\rightarrow 0$ which
allows to solve Eqs.\ (\ref{Iect}) by iteration.

In this letter we have extended previous studies of the ground state
properties of the one dimensional Hubbard model for small interaction
$u$ to the low-lying excitations. Apart from providing the possibility
for a check of perturbative methods our results emphasize the importance
of the ``spinon-holon'' picture for strongly correlated electrons in
particular in the case of a vanishing magnetic field.

This work has been supported in part by the Deutsche
Forschungsgemeinschaft under Grant No.\ Fr~737/2--1.

\newpage

\setlength{\baselineskip}{13pt}

\begin{thebibliography}{10}

\bibitem{liwu:68}
E.~H. Lieb and F.~Y. Wu,
\newblock {\em Phys. Rev. Lett.} {\bf 20} (1968) 1445.

\bibitem{taka:72}
M.~Takahashi,
\newblock {\em Prog. Theor. Phys.} {\bf 47} (1972) 69.

\bibitem{ovch:70}
A.~A. Ovchinnikov,
\newblock {\em Sov. Phys. JETP} {\bf 30} (1970) 1160.

\bibitem{coll:74}
C.~F. Coll, III,
\newblock {\em Phys. Rev. B} {\bf 9} (1974) 2150.

\bibitem{esko:94}
F.~H.~L. E{\ss}ler and V.~E. Korepin,
\newblock {\em Nucl. Phys. B} {\bf 426} (1994) 505.

\bibitem{woec:87b}
F.~Woynarovich and H.-P. Eckle,
\newblock {\em J. Phys. A} {\bf 20} (1987) L443.

\bibitem{woyn:89}
F.~Woynarovich,
\newblock {\em J. Phys. A} {\bf 22} (1989) 4243.

\bibitem{frko:90}
H.~Frahm and V.~E. Korepin,
\newblock {\em Phys. Rev. B} {\bf 42} (1990) 10553.

\bibitem{frko:91}
H.~Frahm and V.~E. Korepin,
\newblock {\em Phys. Rev. B} {\bf 43} (1991) 5653.

\bibitem{taka:71}
M.~Takahashi,
\newblock {\em Progr. Theor. Phys.} {\bf 45} (1971) 756.

\bibitem{ecpo:79}
E.~N. Economou and P.~N. Poulopoulos,
\newblock {\em Phys. Rev. B} {\bf 20} (1979) 4756.

\bibitem{mevo:89}
W.~Metzner and D.~Vollhardt,
\newblock {\em Phys. Rev. B} {\bf 39} (1989) 4462.

\bibitem{yaya:66}
C.~N. Yang and C.~P. Yang,
\newblock {\em Phys. Rev.} {\bf 150} (1966) 327.

\end{thebibliography}

\newpage

\section*{Figures}

\end{document}